\documentclass[final]{nst}

\usepackage{dcolumn}
\usepackage[T2A,T1]{fontenc}
\usepackage[russian,english]{babel}
\usepackage{booktabs,adjustbox}

\usepackage{listings}
\usepackage{color}

\makeatletter
\def\nst@plain@header{}
\def\nst@header{}
\makeatother

\begin{document}

\title{3D Magnetic Field Reconstruction and Mapping with Physics-Informed Neural Networks}

\author{Haohan Yu$^{\circ}$}
\affiliation{State Key Laboratory of Dark Matter Physics, School of Physics and Astronomy, Shanghai Jiao Tong University, Shanghai, China}
\affiliation{Key Laboratory for Particle Astrophysics and Cosmology (Ministry of Education), Shanghai Jiao Tong University, Shanghai, China}
\affiliation{Shanghai Key Laboratory for Particle Physics and Cosmology, Shanghai Jiao Tong University, Shanghai, China}
\author{Zhanxu Hao$^{\circ}$}
\affiliation{State Key Laboratory of Dark Matter Physics, School of Physics and Astronomy, Shanghai Jiao Tong University, Shanghai, China}
\affiliation{Key Laboratory for Particle Astrophysics and Cosmology (Ministry of Education), Shanghai Jiao Tong University, Shanghai, China}
\affiliation{Shanghai Key Laboratory for Particle Physics and Cosmology, Shanghai Jiao Tong University, Shanghai, China}
\author{Bingzhi Li$^{\circ}$}
\email[Corresponding author, ]{bingzhi.li@zhejianglab.org}
\affiliation{Scientific Model Research Group, Zhejiang Lab, Hangzhou, China}
\author{Zejia Lu}
\affiliation{State Key Laboratory of Dark Matter Physics, School of Physics and Astronomy, Shanghai Jiao Tong University, Shanghai, China}
\affiliation{Key Laboratory for Particle Astrophysics and Cosmology (Ministry of Education), Shanghai Jiao Tong University, Shanghai, China}
\affiliation{Shanghai Key Laboratory for Particle Physics and Cosmology, Shanghai Jiao Tong University, Shanghai, China}
\author{Xiang Chen}
\affiliation{State Key Laboratory of Dark Matter Physics, School of Physics and Astronomy, Shanghai Jiao Tong University, Shanghai, China}
\affiliation{Key Laboratory for Particle Astrophysics and Cosmology (Ministry of Education), Shanghai Jiao Tong University, Shanghai, China}
\affiliation{Shanghai Key Laboratory for Particle Physics and Cosmology, Shanghai Jiao Tong University, Shanghai, China}
\author{Liang Li}
\email[Corresponding author, ]{liangliphy@sjtu.edu.cn}
\affiliation{State Key Laboratory of Dark Matter Physics, School of Physics and Astronomy, Shanghai Jiao Tong University, Shanghai, China}
\affiliation{Key Laboratory for Particle Astrophysics and Cosmology (Ministry of Education), Shanghai Jiao Tong University, Shanghai, China}
\affiliation{Shanghai Key Laboratory for Particle Physics and Cosmology, Shanghai Jiao Tong University, Shanghai, China}

\footnotetext{$^{\circ}$ These authors contributed equally to this work.}

\begin{abstract}
Accurate reconstruction of magnetic fields in inaccessible regions is vital for many high-precision experiments in physics. Traditional methods, such as spherical harmonic expansion, often suffer from truncation errors that limit their precision. This study proposes an advanced Physics-Informed Neural Network (PINN) framework for high-precision 3D magnetic field mapping. Unlike conventional data-driven models, the proposed PINN integrates Maxwell's equations directly into the loss function, enforcing divergence-free and curl-free conditions across the entire domain. A key innovation is the inclusion of explicit physics-residual losses at measurement locations, ensuring rigorous physical consistency beyond random collocation sampling. Validation using simulated data achieves a reconstruction accuracy of $10^{-4}$, a tenfold improvement over existing PINN benchmarks.  
Furthermore, experimental validation using a custom coil assembly demonstrates robust reconstruction with sub-percent relative accuracy, reaching the $10^{-3}$ level under ambient conditions.
This AI-driven methodology provides a robust, high-precision solution for field monitoring and measurement in complex experimental environments where direct sensor placement is restricted.
\end{abstract}


\maketitle

\section{Introduction}

The precision of three-dimensional (3D) magnetic field measurement and reconstruction is a cornerstone of many high-precision experiments in physics. A broad class of experimental systems, including charged-particle spectrometers, beamline transport systems, storage-ring measurements, spin-dynamics studies, and magnetic-field-sensitive measurements in condensed matter physics, requires accurate knowledge of 3D magnetic fields across experimentally inaccessible regions. In such systems, magnetic-field maps directly affect trajectory reconstruction, momentum determination, beam matching, spin-precession analyses, and the interpretation of field-dependent material or device responses. Direct probe placement is often limited by vacuum chambers, detector structures, beamline apertures, cryogenic environments, sample holders, or other mechanical constraints. Therefore, reliable reconstruction of magnetic fields from sparse external or boundary measurements is essential for extending field knowledge into regions where direct mapping is impractical. Depending on the experimental context, the required field knowledge may range from sub-percent to ppm or even sub-ppm levels, placing stringent demands not only on the reconstruction algorithm but also on field-source uniformity, probe calibration, temporal stability, and suppression of ambient magnetic noise.

A representative example is the Muon $g-2$/EDM experiment (E34) at J-PARC~\cite{bib:1,bib:2,bib:3}, which requires high-precision magnetic-field knowledge across multiple regions of the experimental apparatus. In the muon beam production and transport region, a well controlled magnetic environment is needed to preserve muon polarization during cooling and to ensure efficient beam delivery. The 3D injection scheme further requires accurate field predictions for beam transport and matching. Finally, within the storage ring, the field must be monitored with sub-ppm stability~\cite{bib:4} to suppress systematic uncertainties in spin-precession measurements. These requirements illustrate a broader challenge faced by high-precision experiments: magnetic fields must often be reconstructed in regions where direct probe placement is obstructed, while the attainable precision depends simultaneously on the reconstruction method, the quality of the magnetic environment, and the control of experimental systematics.

Traditionally, magnetic field reconstruction in source-free regions has relied on solving the Laplace equation, often via Spherical Harmonic Expansion or multipole expansion~\cite{bib:5,bib:6,bib:7}. While these provide a robust mathematical framework, they inherently suffer from ``truncation errors.'' To avoid over-fitting to sensor noise, the expansion series must be terminated at a finite order, which limits the ability to capture high-spatial-frequency variations, especially in regions with high gradients. Furthermore, expansion-based methods are susceptible to noise and require prior knowledge of the specific field profile to regularize expansion terms.

In recent years, machine learning has emerged as a transformative tool across the physical sciences\cite{bib:8,bib:9,bib:10}. Specifically, the paradigm of Physics-Informed Neural Networks (PINNs)\cite{bib:11,bib:12} has emerged as a transformative tool for solving partial differential equations. Unlike traditional ``black-box'' deep learning, PINNs embed the underlying physical laws---specifically Maxwell's equations---directly into the neural network's loss function. A pioneering work~\cite{bib:13} in this field demonstrated the feasibility of using PINNs for magnetic field mapping in inaccessible regions, showing that neural networks can successfully regularize the output to satisfy $\nabla \cdot \mathbf{B} = 0$ and $\nabla \times \mathbf{B} = 0$. However, while this baseline approach proved superior to traditional multipole expansions in noisy environments, its reconstruction accuracy (typically at the $10^{-3}$ level) remains a bottleneck for next-generation precision experiments. This limitation often stems from the reliance on randomly sampled collocation points, which may not sufficiently constrain the model at the boundaries or sensor locations.

This work presents a PINN-based framework specifically optimized for high-precision 3D magnetic field challenges. Our core innovation lies in the introduction of a dual-constrained optimization strategy: in addition to enforcing Maxwell's equations across the entire domain, we implement an explicit physics-residual loss  that applies these physical laws directly at measurement locations. By requiring the network to honor the physics most rigorously where the data is most reliable, we ensure a physically consistent solution that transcends the limitations of random sampling. 

The proposed methodology is validated through both high-fidelity simulations and physical experiments. Simulation results demonstrate a reconstruction accuracy on the order of $10^{-4}$, achieving a tenfold improvement over the baseline PINN approach. 
Furthermore, the model's robustness is verified using data from a custom-built coil assembly, demonstrating sub-percent relative accuracy and reaching the $10^{-3}$ level under ambient conditions. This methodology provides a robust, high-precision foundation for magnetic field mapping in complex environments where traditional probe placement is difficult, and establishes a promising pathway toward the precision requirements of next-generation high-precision experiments.

\section{Simulation and Experimental Setup}

\subsection{Magnetic Field Simulation Framework}

To rigorously evaluate the fundamental performance and convergence properties of the proposed PINN framework, a synthetic magnetic field environment was established based on the Biot-Savart law. The simulation utilizes a configuration of multiple current-carrying circular loops to generate a non-homogeneous magnetic field within a cubic domain. The magnetic induction $\mathbf{B}$ at any observation point $\mathbf{r}$ is analytically calculated by:

\begin{equation}
\mathbf{B}(\mathbf{r}) = \frac{1}{4\pi} \sum_{i} \int \frac{I_i d\mathbf{l}_i \times (\mathbf{r} - \mathbf{r}_i)}{|\mathbf{r} - \mathbf{r}_i|^3}
\end{equation}

The data sampling strategy is designed to mimic a restricted-access experimental scenario and systematically investigate the influence of training data density on reconstruction performance. We perform a parametric sweep by distributing $N_{\text{sample}}$ points uniformly across each of the six boundary faces, with $N_{\text{sample}} \in \{1, 3, 5, 9, 18\}$, yielding sparse training sets ranging from 6 to 108 surface points. To evaluate robustness against measurement uncertainty, zero-mean Gaussian noise with relative standard deviations of $0.0\%$, $0.01\%$, $0.1\%$, and $1.0\%$ (i.e., $\sigma = \alpha |\mathbf{B}|$ with $\alpha$ being the relative noise level) is superimposed on the boundary training data. The interior volume serves as an independent validation set to quantify reconstruction fidelity under the divergence-free and curl-free constraints.

To further validate the framework's applicability to standard field configurations, an additional simulation employing an ideal Helmholtz coil pair was conducted. This configuration generates a highly uniform central field, providing a contrasting test case to the steep gradients of the multi-loop setup. It is important to note that the simulated magnetic field data are treated in a dimensionless framework. The Biot-Savart solution provides the reference field distribution in normalized units, and all reported reconstruction errors for the simulation study are relative quantities (i.e., ratio of absolute error to the local field magnitude). Consequently, the simulation results are independent of the absolute physical field strength and are generally applicable across different current scalings.

\subsection{Experimental Setup and Data Acquisition}

Physical validation was conducted using a custom-engineered 3D magnetic field mapping system integrated with a high-precision automated three-axis translation stage (positioning repeatability $\pm 0.05$~mm). The magnetic field generator utilizes three orthogonal, non-standard coil pairs to create a complex spatial field distribution, purposefully deviating from ideal Helmholtz geometries to evaluate the model's performance in non-uniform gradients:
\begin{itemize}
    \item Circular Coil Pairs: Two sets of circular loops with diameters of 414~mm and 440~mm, with axial spacings of 316~mm and 358~mm, respectively.
    \item Rectangular Coil Pair: A single pair of rectangular coils ($378 \times 272$~mm) with an equivalent spacing of 358~mm.
\end{itemize}

This asymmetric and multi-geometric configuration ensures that the generated field contains sufficient spatial gradients to test the PINN’s reconstruction fidelity beyond simple uniform regions. The coordinate origin ($0,0,0$) was aligned with the geometric center of the coil assembly. To demonstrate the robustness of the PINN framework in unshielded environments, all measurements were conducted under ambient magnetic conditions. 

This configuration allows for the assessment of the framework's performance in reconstructing total local field vectors under ambient conditions, bypassing the requirement for dedicated magnetic shielding. The coil assembly generates a controllable central magnetic field ranging from $0$ to $400~\mu\mathrm{T}$ across the measurement volume, with the maximum field limited by the current capacity of the power supplies.

Magnetic field vectors were measured with an RM3100 magneto-inductive sensor, calibrated using its internal hard-iron and soft-iron compensation routines to ensure measurement fidelity. The sensor is characterized by a resolution of $13~\text{nT}$ and an intrinsic noise floor of $15~\text{nT}$. To suppress transient electronic noise and low-frequency environmental fluctuations, a 10-second dwell time was implemented at each coordinate (10~Hz sampling rate). This acquisition protocol yielded a stable mean measurement with a standard deviation of $\sigma \approx 30$~nT, slightly higher than the intrinsic sensor noise floor of 15~nT due to residual environmental fluctuations. Fig.~\ref{fig:coil} illustrates the design and physical realization of the experimental apparatus. The CAD rendering (left) depicts the integrated measurement system comprising a three-axis translation stage and 3D orthogonal coil assembly. Notably, the yellow non-magnetic fixture secures the RM3100 sensor (indicated in red) for magnetic field measurements, while the green components denote the cable management structure facilitating stage mobility. The photograph (right) presents the physical implementation of this configuration.

Fig.~\ref{fig:exp_data} illustrates the spatial distribution of measurement points for the two experimental sampling strategies employed to evaluate the impact of data quantity on reconstruction accuracy. Fig.~\ref{fig:train_data} presents the dense sampling configuration, utilizing a complete $5\times5\times5$ grid with 125 points spaced at 40~mm intervals throughout the $[-80, 80]$~mm cubic domain. In contrast, Fig.~\ref{fig:train_data_sparse} depicts the sparse sampling strategy, comprising only 30 points restricted to the surface centers (5 points per face). This comparative setup enables direct assessment of PINN reconstruction fidelity under high-data versus severely data-limited conditions.

In the dense configuration, the 98 surface points are used as training data, while the 27 interior points form the held-out evaluation set used in Table~\ref{tab:realdata_summary}.
Throughout the reconstruction volume, all current-carrying coils are located outside the $[-80,80]$ mm cubic domain and no ferromagnetic material is present within it. Over the duration of each scan, the field in this volume is therefore treated as quasi-static and source-free, so that $\nabla\cdot\mathbf{B}=0$ and $\nabla\times\mathbf{B}=0$ apply within the reconstruction domain.

\begin{figure}[!htb]
\centering
\includegraphics[width=0.45\textwidth]{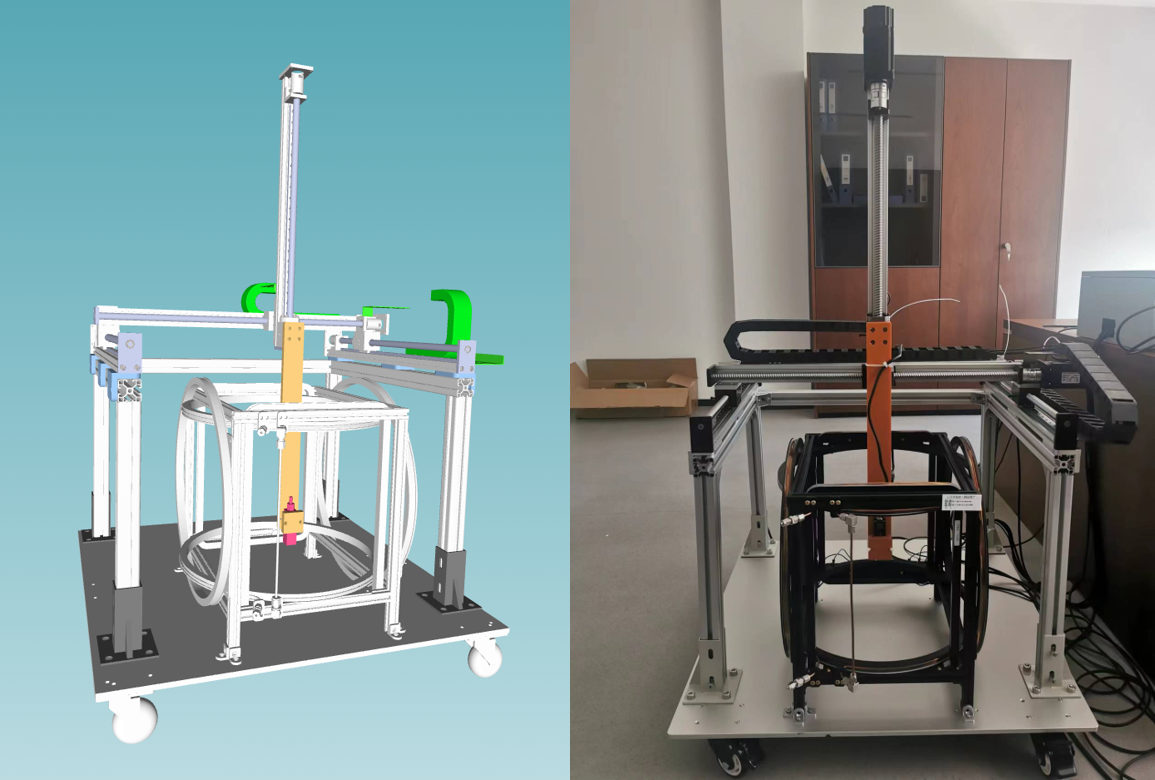}
\caption[CAD rendering and photograph of the 3D magnetic field mapping system]{(Left) CAD rendering and (Right) photograph of the custom 3D magnetic field mapping system, showing the orthogonal coil assembly, three-axis translation stage, and RM3100 sensor fixture in the unshielded laboratory environment.}
\label{fig:coil}
\end{figure}

\begin{figure}[!htb]
\centering
\includegraphics[width=0.45\textwidth]{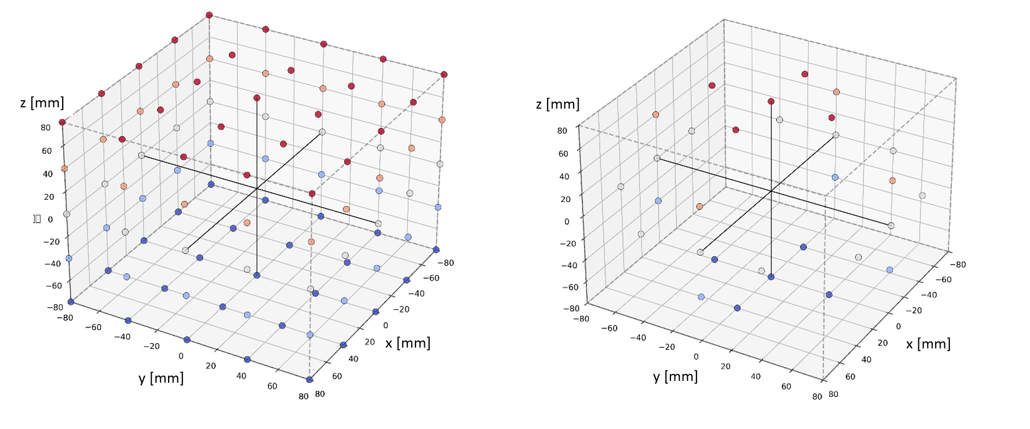}
\caption[]{Spatial sampling strategies for experimental validation. (Left) Dense sampling with a uniform $5\times5\times5$ grid (125 points) over the $[-80, 80]$~mm domain. (Right) Sparse sampling with 30 surface points (5 points per face) restricted to the six boundary surfaces.}
\label{fig:coil}
\end{figure}


\section{Methodology}
\subsection{Laplace multipole expansion}
In source-free regions ($\mathbf{J}=0$) devoid of magnetization ($\mathbf{M}=0$), the reconstruction of the magnetic field is conventionally formulated as a boundary value problem governed by the magnetic scalar potential $\Phi_M$. The magnetic induction $\mathbf{B}$ is derived from the negative gradient of the potential, $\mathbf{B} = -\nabla \Phi_M$, where $\Phi_M$ satisfies the Laplace equation~\cite{bib:7,bib:14}:
\begin{equation}
    \nabla^2 \Phi_M = 0
\end{equation}
In spherical coordinates $(r, \theta, \phi)$, the general solution is typically represented via a multipole expansion utilizing associated Legendre polynomials $P_l^m$:
\begin{multline}
    \Phi_M(r, \theta, \phi) = \sum_{l=1}^{\infty} \sum_{m=0}^{l} r^l P_l^m(\cos \theta) \\
    \times [a_{lm} \cos(m\phi) + b_{lm} \sin(m\phi)]
\end{multline}
where $a_{lm}$ and $b_{lm}$ are the expansion coefficients determined by boundary conditions. In many precision experiments, the field is often decomposed into a compact basis: 
$\mathbf{B}(x,y,z) = \sum_{n} c_n \mathbf{f}_n(x,y,z)$, where $\mathbf{f}_n$ denotes the basis vector functions. Despite its mathematical elegance, this approach is fundamentally constrained by truncation errors. To maintain numerical stability against experimental noise, the expansion must be terminated at a finite order. Consequently, this truncation restricts the ability to resolve high-spatial-frequency variations, which are critical in complex detector geometries or regions with steep gradients.

To ensure a rigorous and fair comparison against the proposed PINN framework, the multipole expansion order was systematically optimized for each experimental configuration. We performed a parametric scan of truncation orders up to 12, selecting the order that minimized the validation error without exhibiting overfitting to measurement noise. 
This adaptive selection strategy provides a conservative best-case benchmark for comparison with the PINN; because it uses interior evaluation points, it is not intended to represent an operational reconstruction protocol for a truly inaccessible region.

\subsection{Proposed PINN framework}

To circumvent the resolution bottlenecks inherent in expansion-based methods, we propose an advanced physics-informed neural network (PINN) framework. The network, denoted as $f_{\theta}$, functions as a continuous and differentiable approximator that maps normalized spatial coordinates $(x, y, z)$ to the magnetic field vector $\mathbf{B} = (B_x, B_y, B_z)$.

\paragraph{Network Architecture}
We adopt a deep fully connected architecture utilizing Sinusoidal Representation Networks (SIREN)~\cite{bib:15,bib:16} with periodic sine activation functions. Unlike traditional ReLU or Tanh activations, the periodic nature of the sine function enhances the network's capacity to represent complex oscillatory and high-frequency components characteristic of magnetic fields. To facilitate gradient flow and enable efficient training of deep architectures, residual skip connections are incorporated, aggregating outputs from preceding layers to preserve multi-scale feature information. The inherent differentiability of the sine function allows for the computation of exact spatial derivatives through automatic differentiation, ensuring that Maxwellian constraints are satisfied with machine precision without numerical discretization errors.

\paragraph{Data Preprocessing}
To ensure numerical stability and accelerate convergence, input coordinates and magnetic field labels undergo rigorous normalization. Spatial coordinates are scaled to the range $[-1, 1]$ via coordinate-wise division by the domain half-length $L$. Magnetic field components are standardized using z-score normalization: $\tilde{\mathbf{B}} = (\mathbf{B} - \boldsymbol{\mu}) / \boldsymbol{\sigma}$, where $\boldsymbol{\mu}$ and $\boldsymbol{\sigma}$ denote the mean and standard deviation computed from the training set. During inference, predictions are inverse-transformed via $\mathbf{B}_{\text{pred}} = \tilde{\mathbf{B}}_{\text{NN}} \times \boldsymbol{\sigma} + \boldsymbol{\mu}$ to restore physical units. This preprocessing mitigates gradient vanishing issues and balances the multi-component loss terms during optimization.
When evaluating the Maxwell losses, the derivatives are expressed in physical units: the chain-rule factors associated with coordinate normalization and field standardization are included before forming $\nabla\cdot\mathbf{B}$ and $\nabla\times\mathbf{B}$.

\paragraph{Composite Loss Function}
The optimization objective is governed by a composite loss function $\mathcal{L}$ that synergizes empirical data with rigorous physical laws. A pivotal innovation in this framework is the dual-enforcement of physics constraints: they are applied not only to random collocation points across the domain but also explicitly at the sensor coordinates to ensure local physical consistency.

The multi-component data loss ($\mathcal{L}_{\text{data}}$) guarantees both directional fidelity and magnitude precision:
\begin{equation}
\begin{split}
\mathcal{L}_{\text{data}} = & \frac{1}{N_B} \sum_{i=1}^{N_B} \|\mathbf{B}_{\text{pred}}(\mathbf{r}_d^i) - \mathbf{B}_{\text{gt}}(\mathbf{r}_d^i)\|^2 \\
& + \frac{\lambda_{m}}{N_B} \sum_{i=1}^{N_B} \left| \|\mathbf{B}_{\text{pred}}(\mathbf{r}_d^i)\| - \|\mathbf{B}_{\text{gt}}(\mathbf{r}_d^i)\| \right|^2,
\end{split}
\end{equation}
where $\lambda_{m}$ is the weighting coefficient for the field magnitude constraint, $\mathbf{B}_{\text{gt}}$ is the ground truth magnetic field. In practice, we set $\lambda_{m} = 1.0$ to balance directional and magnitude fidelity.

The explicit physics-informed loss ($\mathcal{L}_{\text{phys}}$) enforces the fundamental Maxwell's equations for divergence-free ($\nabla \cdot \mathbf{B} = 0$) and curl-free ($\nabla \times \mathbf{B} = 0$) conditions throughout the source-free region:
\begin{equation}
\begin{split}
\mathcal{L}_{\text{phys}} = & \frac{1}{N} \sum_{j \in \{\mathbf{r}_f, \mathbf{r}_d\}} \|\nabla \cdot \mathbf{B}_{\text{pred}}(\mathbf{r}^j)\|^2 \\
& + \frac{1}{N} \sum_{j \in \{\mathbf{r}_f, \mathbf{r}_d\}} \|\nabla \times \mathbf{B}_{\text{pred}}(\mathbf{r}^j)\|^2.
\end{split}
\end{equation}
Crucially, these laws are evaluated at both the internal collocation points $\mathbf{r}_f$ and the boundary detector positions $\mathbf{r}_d$, ensuring that physical laws are most rigorously enforced where measurement data is available. Here, $N = N_f + N_d$ denotes the total number of collocation and detector points. To facilitate explicit monitoring of constraint satisfaction at measurement locations, the divergence and curl components of $\mathcal{L}_{\text{phys}}$ evaluated exclusively at the boundary sensor coordinates $\mathbf{r}_d$ are defined as
\begin{equation}
\mathcal{L}_{\text{BC div}} = \frac{1}{N_d} \sum_{j \in \mathbf{r}_d} \|\nabla \cdot \mathbf{B}_{\text{pred}}(\mathbf{r}^j)\|^2
\end{equation}
and
\begin{equation}
\mathcal{L}_{\text{BC curl}} = \frac{1}{N_d} \sum_{j \in \mathbf{r}_d} \|\nabla \times \mathbf{B}_{\text{pred}}(\mathbf{r}^j)\|^2,
\end{equation}
respectively, where $N_d$ is the number of boundary detector points. These physics residuals at measurement locations enable direct assessment of Maxwellian constraint adherence at the sensor positions during training.

\paragraph{Three-Stage Training Strategy}
The framework employs a progressive three-stage optimization pipeline to achieve high-precision convergence and robustness against measurement noise:

\textbf{Stage 1: AdamW Pre-training.} The network is initially optimized using the AdamW algorithm~\cite{bib:17} with decoupled weight decay. This first-order method rapidly navigates the optimization landscape to the neighborhood of a global minimum while mitigating overfitting to sensor noise.

\textbf{Stage 2: LBFGS Fine-tuning.} Upon convergence of Stage 1, the optimizer switches to Limited-memory Broyden-Fletcher-Goldfarb-Shanno (LBFGS)~\cite{bib:18} with strong Wolfe line search. This second-order quasi-Newton method exploits local curvature information to achieve higher precision, refining the solution to high precision.

\textbf{Stage 3: Residual-based Adaptive Refinement (RAR).} To address potential residual concentrations in under-sampled regions, an adaptive refinement stage is implemented~\cite{bib:19}. 
In each RAR iteration, point selection is based on the divergence residual $R=|\nabla\cdot\mathbf B|$, which exhibited the dominant localized violations in the present tests, while both divergence and curl residuals are retained in the augmented loss. Points exhibiting maximal residual magnitudes are identified and added to an expanding set of extra collocation points $\mathbf{r}_{\text{rar}}$. The loss function is augmented with an additional term:
\begin{equation}
\begin{split}
\mathcal{L}_{\text{rar}} = & \frac{1}{N_{\text{rar}}} \sum_{j \in \mathbf{r}_{\text{rar}}} \|\nabla \cdot \mathbf{B}_{\text{pred}}(\mathbf{r}^j)\|^2 \\
& + \frac{1}{N_{\text{rar}}} \sum_{j \in \mathbf{r}_{\text{rar}}} \|\nabla \times \mathbf{B}_{\text{pred}}(\mathbf{r}^j)\|^2.
\end{split}
\end{equation}
This adaptive sampling strategy dynamically concentrates computational resources on regions where Maxwell's equations are most violated, ensuring uniform physical consistency across the entire domain.

The framework is implemented using the PyTorch library, leveraging GPU acceleration for automatic differentiation and parallel computation of physics constraints across collocation points. The methodology enables continuous spatial reconstruction without grid-based discretization, facilitating arbitrary-resolution magnetic field mapping.

\section{Results and Discussion}

\subsection{Simulation Validation}

\begin{figure}[!htb]
\centering
\includegraphics[width=0.45\textwidth]{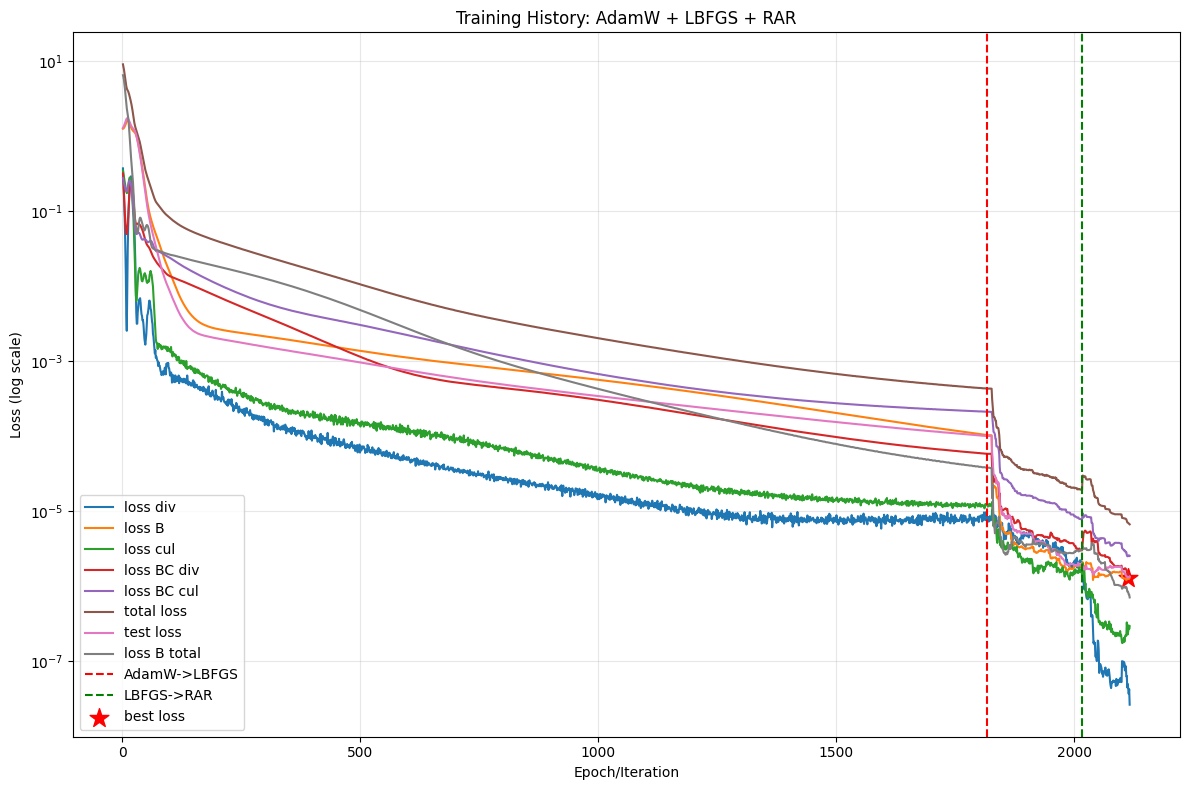}
\caption{Training convergence history for the ideal synthetic data, illustrating the three-stage optimization process (AdamW $\rightarrow$ LBFGS $\rightarrow$ RAR). Vertical dashed lines mark stage transitions, and stars indicate the epochs with the minimal monitored test loss.}
\label{fig:loss}
\end{figure}

\paragraph{Training Convergence and Optimization Dynamics}
The proposed three-stage training strategy demonstrates distinct optimization dynamics on synthetic Biot-Savart data, as illustrated in Fig.~\ref{fig:loss}. During Stage 1 (AdamW pre-training), all loss components exhibit rapid convergence, with the data loss ($\mathcal{L}_{\text{data}}$) and physics-residual losses at measurement locations ($\mathcal{L}_{\text{BC div}}$, $\mathcal{L}_{\text{BC curl}}$) decreasing by approximately two orders of magnitude. Notably, the explicit enforcement of physics constraints at measurement locations ($\mathcal{L}_{\text{BC div}}$ and $\mathcal{L}_{\text{BC curl}}$, representing the divergence and curl components of $\mathcal{L}_{\text{phys}}$ evaluated specifically at boundary sensor coordinates $\mathbf{r}_d$) remains consistently lower than the domain-averaged physics losses, validating the efficacy of the 
dual-constraint strategy. The transition to Stage 2 (LBFGS) yields immediate refinement, particularly evident in the divergence loss ($\mathcal{L}_{\text{div}}$) dropping below $10^{-5}$. 

The most significant enhancement occurs during Stage 3 (RAR), where adaptive resampling of high-residual regions drives the total loss below $10^{-7}$. The test loss closely tracks the training total loss without significant divergence, indicating robust generalization to unseen interior regions.
The interior evaluation points are used only for post-training assessment and are not included in the loss function, early stopping, or model selection.

\begin{figure}[!htb]
\centering
\includegraphics[width=0.45\textwidth]{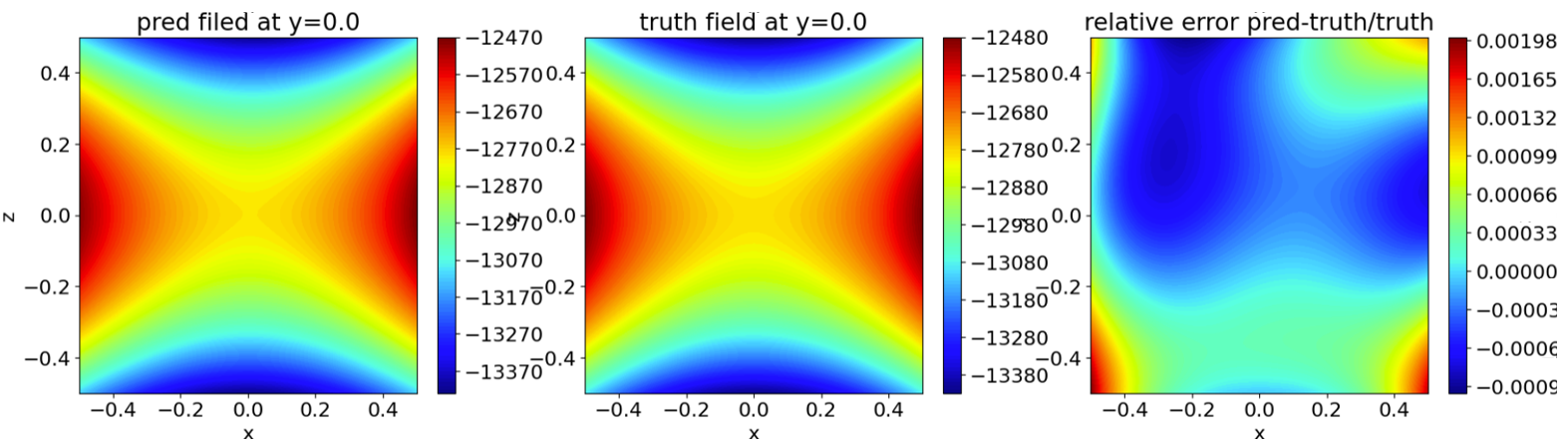}
\caption{Spatial reconstruction quality at the $y=0$ section for simulated data with sparse boundary sampling ($N_{\text{sample}}=3$ per face, noise-free). The panels show (left) PINN prediction, (middle) Biot-Savart ground truth, and (right) relative error distribution, demonstrating $\sim10^{-4}$ relative error despite severely limited training data.}
\label{fig:field_slices}
\end{figure}

\paragraph{Reconstruction Accuracy and Spatial Resolution}
Validation against the analytical Biot-Savart solution reveals exceptional reconstruction fidelity across the entire cubic domain. Fig.~\ref{fig:field_slices} presents comparative visualizations at the representative $y=0$-sections, obtained with only $N_{\text{sample}}=3$ boundary points per face (18 total surface measurements), demonstrating the PINN's capability to capture complex non-uniform field patterns generated by the multi-loop configuration from severely sparse observations.

Quantitative assessment indicates relative errors below $4 \times 10^{-3}$ throughout the domain, with typical values ranging from $10^{-4}$ to $10^{-3}$. The achieved reconstruction accuracy of $\sim 10^{-4}$ represents a tenfold improvement over existing PINN-based methods, which typically report errors at the $10^{-3}$ level. This enhancement stems primarily from the explicit physics residual enforcement at measurement locations and the RAR refinement strategy.

Compared to spherical harmonic expansion methods, the PINN approach eliminates truncation artifacts that manifest as Gibbs oscillations near steep gradients. Furthermore, unlike expansion methods that require regularization to prevent overfitting to noise, the physics-informed constraints provide natural regularization without manual tuning of expansion orders. The continuous functional representation enables arbitrary spatial resolution without interpolation errors, providing a significant advantage for high-gradient regions in constrained high-precision experimental systems.

Remarkably, the PINN framework achieves consistent reconstruction accuracy ($\sim$$10^{-4}$) across both the complex multi-loop configuration and the ideal Helmholtz coil geometry. This insensitivity to field topology validates that the dual-constraint strategy generalizes across varying spatial gradient characteristics, from highly non-uniform to near-uniform field distributions.

\begin{figure}[!htb]
\includegraphics[width=0.45\textwidth]{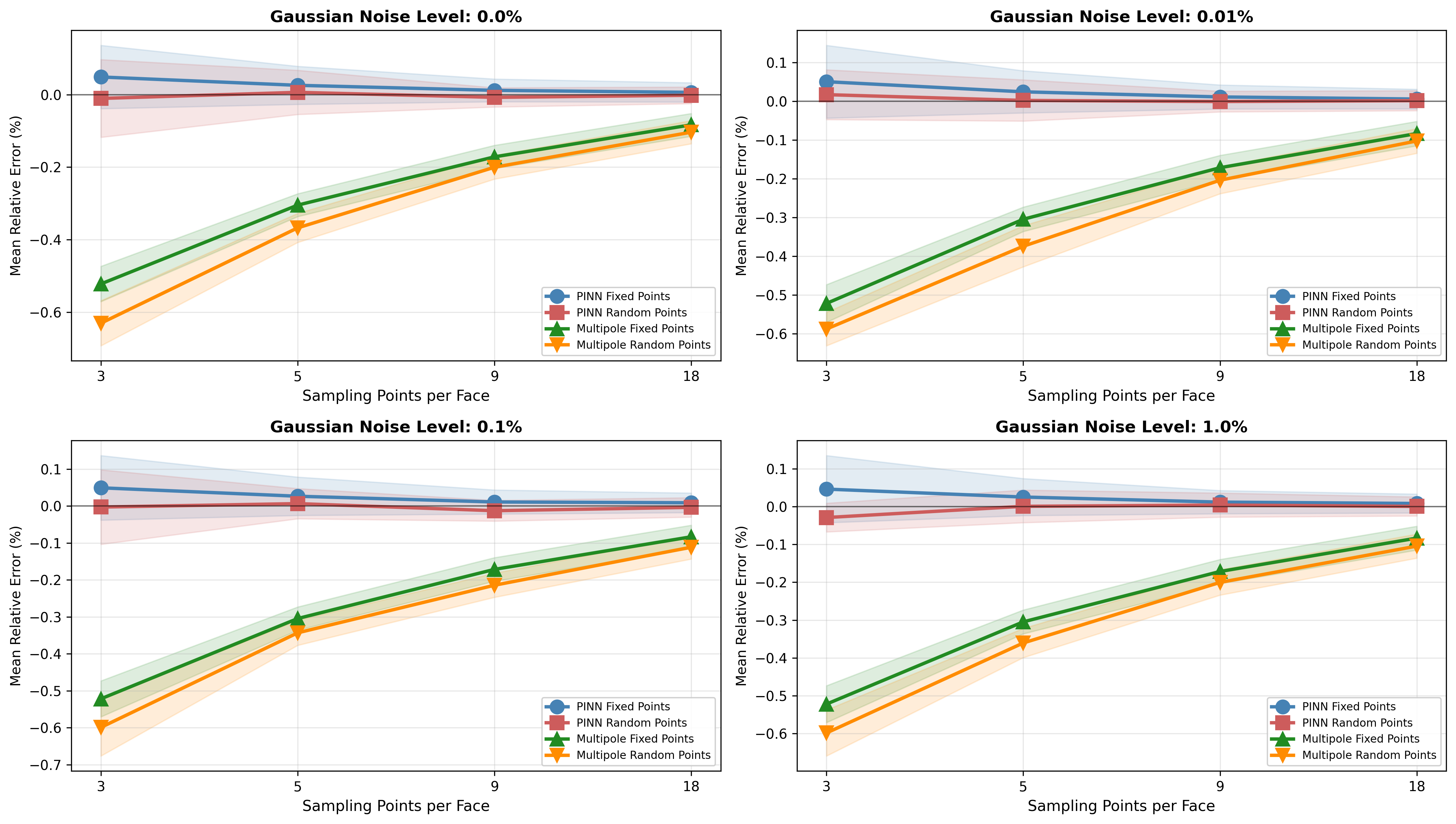}
\caption{Field prediction error vs. sampling density under Gaussian noise (0--1\%). PINN (blue: fixed, red: random) maintains $\sim$0.01\% mean signed relative error with weak dependence on noise and sampling, while multipole expansion (green: fixed, orange: random) exhibits much larger systematic deviations that decrease with denser sampling.}
\label{fig:error}
\end{figure}

\begin{table*}[htbp]
\centering
\caption{Robustness analysis of magnetic-field prediction under varying Gaussian noise levels. Mean signed relative error is shown versus sampling points per face for both PINN and multipole expansion methods.}
\label{tab:helmholtz_sci}
\begin{adjustbox}{width=\textwidth,center}
\begin{tabular}{c|c|cc|cc|cc|cc|cc}
\toprule
Noise (\%) & Method & \multicolumn{2}{c|}{$N_{sample}=1$} & \multicolumn{2}{c|}{$N_{sample}=3$} & \multicolumn{2}{c|}{$N_{sample}=5$} & \multicolumn{2}{c|}{$N_{sample}=9$} & \multicolumn{2}{c}{$N_{sample}=18$} \\
 &  & Fixed & Random & Fixed & Random & Fixed & Random & Fixed & Random & Fixed & Random \\
\midrule
0.0 & PINN & $-1.5 \times 10^{-4}$ & $-2.6 \times 10^{-4}$ & $4.9 \times 10^{-4}$ & $-1.1 \times 10^{-4}$ & $2.6 \times 10^{-4}$ & $6.2 \times 10^{-5}$ & $1.2 \times 10^{-4}$ & $-7.6 \times 10^{-5}$ & $6.2 \times 10^{-5}$ & $-1.9 \times 10^{-5}$ \\
 & Multipole & $-1.6 \times 10^{-2}$ & $-2.1 \times 10^{-2}$ & $-5.2 \times 10^{-3}$ & $-6.3 \times 10^{-3}$ & $-3.0 \times 10^{-3}$ & $-3.7 \times 10^{-3}$ & $-1.7 \times 10^{-3}$ & $-2.0 \times 10^{-3}$ & $-8.3 \times 10^{-4}$ & $-1.0 \times 10^{-3}$ \\
\midrule
0.01 & PINN & $4.2 \times 10^{-4}$ & $1.3 \times 10^{-3}$ & $5.0 \times 10^{-4}$ & $1.7 \times 10^{-4}$ & $2.4 \times 10^{-4}$ & $2.0 \times 10^{-5}$ & $1.1 \times 10^{-4}$ & $-7.6 \times 10^{-6}$ & $5.6 \times 10^{-5}$ & $1.5 \times 10^{-5}$ \\
 & Multipole & $-1.6 \times 10^{-2}$ & $-1.7 \times 10^{-2}$ & $-5.2 \times 10^{-3}$ & $-5.9 \times 10^{-3}$ & $-3.0 \times 10^{-3}$ & $-3.7 \times 10^{-3}$ & $-1.7 \times 10^{-3}$ & $-2.0 \times 10^{-3}$ & $-8.3 \times 10^{-4}$ & $-1.0 \times 10^{-3}$ \\
\midrule
0.1 & PINN & $5.5 \times 10^{-4}$ & $-7.8 \times 10^{-4}$ & $4.9 \times 10^{-4}$ & $-2.9 \times 10^{-5}$ & $2.6 \times 10^{-4}$ & $6.4 \times 10^{-5}$ & $1.1 \times 10^{-4}$ & $-1.3 \times 10^{-4}$ & $8.7 \times 10^{-5}$ & $-3.7 \times 10^{-5}$ \\
 & Multipole & $-1.6 \times 10^{-2}$ & $-1.6 \times 10^{-2}$ & $-5.2 \times 10^{-3}$ & $-6.0 \times 10^{-3}$ & $-3.0 \times 10^{-3}$ & $-3.4 \times 10^{-3}$ & $-1.7 \times 10^{-3}$ & $-2.1 \times 10^{-3}$ & $-8.3 \times 10^{-4}$ & $-1.1 \times 10^{-3}$ \\
\midrule
1.0 & PINN & $4.1 \times 10^{-4}$ & $-7.2 \times 10^{-4}$ & $4.6 \times 10^{-4}$ & $-2.9 \times 10^{-4}$ & $2.5 \times 10^{-4}$ & $5.5 \times 10^{-6}$ & $1.2 \times 10^{-4}$ & $3.9 \times 10^{-5}$ & $8.1 \times 10^{-5}$ & $3.6 \times 10^{-6}$ \\
 & Multipole & $-1.6 \times 10^{-2}$ & $-2.3 \times 10^{-2}$ & $-5.2 \times 10^{-3}$ & $-6.0 \times 10^{-3}$ & $-3.0 \times 10^{-3}$ & $-3.6 \times 10^{-3}$ & $-1.7 \times 10^{-3}$ & $-2.0 \times 10^{-3}$ & $-8.3 \times 10^{-4}$ & $-1.0 \times 10^{-3}$ \\
\bottomrule
\end{tabular}
\end{adjustbox}
\end{table*}

\paragraph{Noise Resilience and Sampling Efficiency}
To evaluate practical applicability under experimental conditions, we systematically investigate the method's robustness against measurement noise and sampling sparsity.
Table~\ref{tab:helmholtz_sci} presents the relative error statistics across four noise levels (0.0\% to 1.0\%) and varying sampling densities. The Gaussian noise is modeled as $N(0, (\sigma B)^2)$, where the standard deviation scales proportionally with the local field magnitude $B$, emulating realistic sensor noise characteristics.

Two sampling strategies are employed to assess geometric bias: \textit{fixed} sampling places measurement points at symmetric positions on each face (e.g., centered or uniformly distributed grid points), whereas \textit{random} sampling draws points uniformly at random across each face surface. The fixed strategy emulates standardized measurement protocols with deterministic probe placement, while the random strategy tests sensitivity to arbitrary probe positioning common in automated scanning systems.

The PINN demonstrates remarkable noise resilience with the magnitude of the mean signed relative errors remaining below $10^{-3}$ for moderate sampling densities ($N_{\rm sample}\ge 3$), even at the 1.0\% injected-noise level where traditional multipole expansions exhibit errors exceeding $5 \times 10^{-3}$ at coarse sampling ($N_{\text{sample}}=3$). Notably, PINN performance exhibits weak dependence on sampling density; increasing sampling points from 3 to 18 per face yields marginal improvement in PINN accuracy (error reduction $< 2\times10^{-4}$), whereas multipole methods show substantial error reduction (approximately 50\% decrease) over the same range. This insensitivity to sampling density stems from the physics-informed regularization, which effectively interpolates between sparse measurements while rigorously satisfying Maxwell's equations. 

Under extremely sparse conditions ($N_{\text{sample}}=1$), the PINN exhibits negligible systematic bias regardless of whether fixed or random sampling is employed, whereas multipole methods show significant systematic errors that only diminish with increased sampling density. 
In the tested random sampling realization, PINN performance remains comparable to that obtained with fixed sampling (differences $< 10^{-4}$), suggesting reduced sensitivity to probe-placement geometry. In contrast, the multipole method exhibits somewhat larger deviations under random sampling at low densities, attributable to the sensitivity of harmonic expansion coefficients to probe placement geometry. Moreover, multipole accuracy depends sensitively on truncation order selection—optimal performance requires case-specific tuning, demanding denser sampling to determine appropriate orders and reduce systematic deviations.

Statistical analysis over $10^6$ spatial points confirms that PINN error distributions remain tightly bounded (narrow shaded regions indicate low standard deviations) across all configurations, whereas multipole methods display broader variance, particularly at low sampling rates. The results validate that explicit enforcement of physical constraints provides inherent regularization against overfitting to noisy data, eliminating the need for manual tuning of expansion orders required by harmonic expansion approaches.

\paragraph{Physical Consistency and Maxwellian Constraints}
A critical advantage of the framework lies in the rigorous satisfaction of Maxwell's equations throughout the computational domain. The divergence-free and curl-free constraints achieve numerical satisfaction at the level of $10^{-5}$ to $10^{-7}$, significantly exceeding the reconstruction accuracy of the field components themselves. 


\begin{figure}[!htb]
\includegraphics[width=0.45\textwidth]{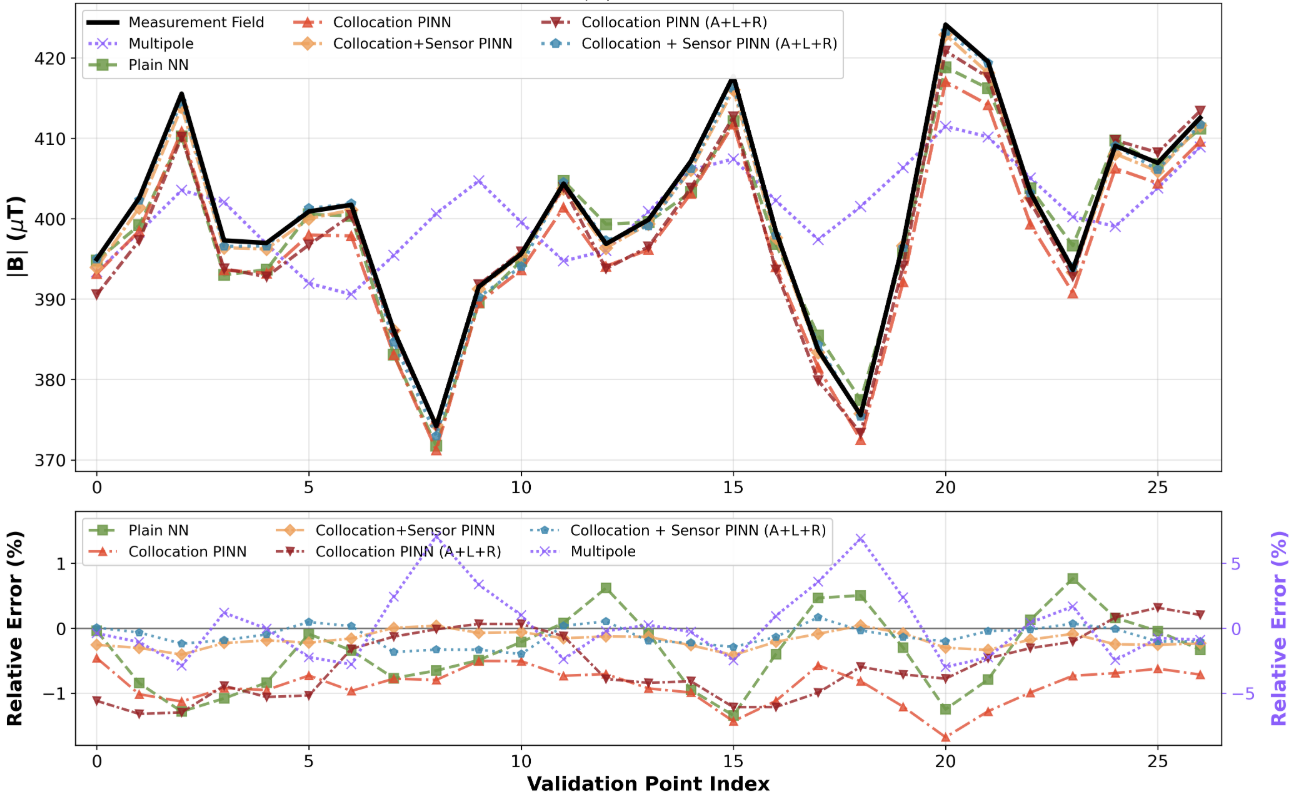}
\caption{Comparison of reconstructed magnetic-field magnitudes $|B|$ at the validation points for $B_0=400~\mu\mathrm{T}$.  The upper panel compares the measured field with the multipole expansion, the plain data-only neural network, and ablation variants of the PINN framework. The lower panel shows the corresponding relative errors with respect to the measurement field.}
\label{fig:exp_reconstruction}
\end{figure}


\begin{table*}[htbp]
\centering
\setlength{\tabcolsep}{7pt}
\caption{Performance comparison of PINN and multipole methods: absolute error statistics ($\mu$T) on experimental magnetic field data across varying central field strengths ($B_0=0$--$400~\mu\mathrm{T}$) and training point densities (30 and 98 points).}
\label{tab:realdata_summary}
\begin{tabular}{c|c|c|c|c|c|c}
\toprule
Center $B_0$ ($\mu$T) & $|B|$ Range ($\mu$T) & Training Points & \multicolumn{2}{c|}{PINN (Mean$\pm$Std)} & \multicolumn{2}{c}{Multipole (Mean$\pm$Std)} \\
 &  &  & Abs Err ($\mu$T) & Rel Err (\%) & Abs Err ($\mu$T) & Rel Err (\%) \\
\midrule
0 & 0.123--3.266 & 30 & 0.201$\pm$0.165 & 19.016$\pm$40.047 & 1.336$\pm$0.770 & 77.742$\pm$69.282 \\
0 & 0.123--3.266 & 98 & 0.219$\pm$0.151 & 21.694$\pm$38.696 & 0.973$\pm$0.534 & 103.925$\pm$217.398 \\
10 & 7.478--12.400 & 30 & 0.145$\pm$0.129 & 1.452$\pm$1.270 & 0.875$\pm$0.707 & 9.212$\pm$8.410 \\
10 & 7.478--12.400 & 98 & 0.067$\pm$0.055 & 0.679$\pm$0.569 & 1.112$\pm$0.663 & 10.973$\pm$6.159 \\
100 & 94.321--107.824 & 30 & 0.197$\pm$0.140 & 0.194$\pm$0.132 & 2.136$\pm$1.710 & 2.116$\pm$1.670 \\
100 & 94.321--107.824 & 98 & 0.134$\pm$0.082 & 0.132$\pm$0.078 & 2.201$\pm$1.428 & 2.201$\pm$1.444 \\
400 & 374.210--424.142 & 30 & 1.679$\pm$0.805 & 0.416$\pm$0.189 & 9.332$\pm$6.298 & 2.328$\pm$1.587 \\
400 & 374.210--424.142 & 98 & 0.935$\pm$0.419 & 0.231$\pm$0.098 & 8.099$\pm$6.614 & 2.046$\pm$1.743 \\
\bottomrule
\end{tabular}
\end{table*}

\subsection{Experimental Validation}

\paragraph{Training Dynamics Under Experimental Conditions}
The efficacy of the framework is further validated using experimental data from the custom-built coil assembly generating a central field of $400~\mu$T. The training dynamics on real-world data exhibit characteristics similar to those observed in the idealized simulation, but with distinct differences arising from experimental uncertainties. While the three-stage optimization pipeline (AdamW $\rightarrow$ LBFGS $\rightarrow$ RAR) maintains robust convergence, the data loss ($\mathcal{L}_{\text{data}}$) converges to a higher floor ($\sim 10^{-4}$) due to the inherent sensor noise ($\sigma \approx 30$~nT, characterized by 10-second averaging) and ambient field fluctuations, contrasting with the machine-precision convergence observed in synthetic data. 

Nevertheless, the RAR stage successfully refines the divergence loss to below $10^{-6}$ despite experimental uncertainties. This demonstrates the framework's ability to distinguish between physical field variations and measurement noise through the Maxwellian constraints.

\paragraph{Reconstruction Precision}
Figure~\ref{fig:exp_reconstruction} compares the reconstructed magnetic-field magnitudes $|B|$ at the 27 interior validation points of the $3\times3\times3$ interior grid, a subset of the full $5\times5\times5$ measurement array. Both the PINN and multipole approaches capture the overall spatial variation of the field induced by the non-ideal coil geometry. However, the PINN predictions show systematically better agreement with the target values across the full validation set. This improvement is evident in the lower panel, where the PINN relative errors remain small and uniform, whereas the multipole reconstruction exhibits noticeably larger point-to-point fluctuations. 

To isolate the effect of physics-informed regularization, we trained a data-only neural network, hereafter referred to as the plain NN, using the same network architecture, input/output normalization, and boundary training data as the PINN, but optimized with $L_{\rm data}$ only, with all Maxwell-equation-based loss terms omitted. The AdamW and LBFGS stages were kept identical to those of the PINN, whereas the RAR stage was omitted because it is defined through physics residuals.
Although the plain NN achieves a mean signed relative error comparable to that of the PINN ($\mu = 0.313\%$ versus $-0.231\%$), its standard deviation is approximately three times larger ($\sigma = 0.321\%$ versus $0.098\%$). Consequently, the plain NN predictions display markedly larger spatial fluctuations and lack the stability exhibited by the PINN, underscoring that the embedded physical constraints are essential for suppressing error variance and ensuring uniformly accurate reconstruction.

An ablation study further dissects the contribution of each component, as shown in Figure~\ref{fig:exp_reconstruction}. The baseline collocation PINN---trained with boundary data and interior PDE residuals---already suppresses the large deviations seen in the multipole reconstruction, keeping relative errors within roughly $\pm 1\%$. Equipping the same architecture with the full A+L+R optimization schedule (AdamW pre-training, LBFGS refinement, and RAR adaptive residual sampling) yields a modest further improvement, smoothing the error profile and reducing point-to-point scatter. The most substantial gain comes from incorporating the interior sensor measurements: the collocation+sensor PINN closely tracks the sharp field extrema near validation points 3, 15, and 20 that the sensor-free variants capture less faithfully. Finally, the complete configuration, fusing collocation points, sensor data, and the A+L+R three-stage optimization, achieves the tightest agreement with the measurement field, exhibiting the smallest error magnitude and the least spatial variation across the entire validation set. This progression confirms that the synergy of sparse interior measurements, embedded Maxwell constraints, and the staged A+L+R optimization is essential for uniformly accurate reconstruction.

Quantitatively, at the 400~$\mu$T operating point, the PINN yields mean relative errors of 0.231\% with 98 training points and 0.416\% with 30 training points, substantially smaller than the corresponding multipole errors.
Notably, this accuracy is maintained even when the boundary training data are reduced to only 30 surface points, corresponding to five points per face, indicating that the PINN reconstruction remains robust under sparse boundary sampling.

\paragraph{Systematic Error Analysis and Method Comparison}
Table~\ref{tab:realdata_summary} provides a systematic quantification of the reconstruction performance across the full operational dynamic range ($0$--$400~\mu$T) and varying data densities. 
For $B_0\le 100~\mu\mathrm{T}$, the PINN achieves sub-$\mu\mathrm{T}$ mean absolute error. At $B_0=400~\mu\mathrm{T}$, the mean absolute error is at $\mu\mathrm{T}$ level, while the relative error remains below 0.5\%.
Notably, this precision is preserved even when the field strength increases from $0$ to $400~\mu$T, demonstrating robust scale-invariant performance when evaluated by the physically appropriate metric.

Several characteristics of the error statistics warrant specific discussion. Beyond the pointwise statistical sensor noise, the present validation may also be affected by residual probe-calibration errors, axis misalignment, stage-positioning uncertainty coupled to field gradients, coil-current stability, and temporal drift of the ambient field during the scan. These effects are not separately disentangled in the present study and should be regarded as part of the experimental uncertainty of the validation setup.
At $B_0=0$, the central field is nulled by adjusting the current ratios of the three orthogonal coil pairs to cancel the geomagnetic field and any residual background, rather than by de-energizing the coils. The resulting field distribution spans $|B| \in [0.123, 3.266]$~$\mu$T, representing the residual gradient field generated by the coil configuration itself. Near the geometric center where $|B| \to 0$, the relative error $\delta B/|B|$ becomes ill-defined due to the vanishing denominator, resulting in the large reported variance ($\sim$40\%) that reflects a mathematical singularity rather than reconstruction instability. In this regime, the absolute error of $\sim$0.2~$\mu$T---comparable to the sensor noise floor---is the valid metric, confirming that the PINN correctly reconstructs the nulled field distribution generated by the active coil system.

At $B_0=400$~$\mu$T, the absolute error increases to $0.94$--$1.68$~$\mu$T (depending on sampling density), yet the relative error remains excellent at $0.23$--$0.42$\%. This reflects the expected linear scaling behavior: larger coil currents generate steeper spatial gradients, evidenced by the increase in the field span ($|B|_{\max} - |B|_{\min}$) from approximately $13.5~\mu$T at $100~\mu$T to $49.9~\mu$T at $400~\mu$T across the measurement volume. Consequently, the absolute residuals scale proportionally with the field magnitude while the fractional fidelity is maintained. Notably, the accuracy improvement from 30 to 98 training points is more pronounced at 400~$\mu$T ($\sim$44\% error reduction) than at lower fields, indicating that high-gradient regions benefit more from increased boundary constraints. Across all non-zero field strengths ($B_0 \geq 10$~$\mu$T), the PINN maintains sub-1\% relative error, substantially outperforming traditional methods.

In contrast, the multipole expansion method exhibits significantly degraded accuracy, with mean absolute errors ranging from $0.67~\mu$T to $9.33~\mu$T---approximately one order of magnitude worse than PINN---and substantially larger variances indicating poor noise resilience. The performance gap widens dramatically at higher field strengths: at $B_0 = 400~\mu$T, the multipole error exceeds $8~\mu$T whereas the PINN achieves a sub-$\mu\mathrm{T}$ mean absolute error in the 98-point training case ($0.935 \pm 0.419~\mu\mathrm{T}$).

Regarding data efficiency, reducing the training points from $98$ to $30$ (a $69\%$ reduction) results in minimal accuracy degradation for PINN---mean error increases by only $0.05$--$0.8~\mu$T depending on field strength, with the most negligible impact observed at low fields ($< 0.02~\mu$T increase at $B_0 = 10~\mu$T). This contrasts sharply with the multipole method, which shows irregular and unpredictable performance variations under reduced sampling. Furthermore, the consistently lower standard deviation in PINN predictions ($\sigma_{\text{PINN}} \ll \sigma_{\text{multipole}}$) confirms that the Maxwellian constraints function as an effective physics-informed regularization mechanism, distinguishing true field variations from stochastic measurement noise.

\paragraph{Implications for High-Precision Experiments}
The experimental validation confirms that the dual-constraint strategy effectively handles sparse, noisy measurements while maintaining physical consistency across the tested dynamic range. The demonstrated sub-percent relative accuracy is maintained even with substantially reduced sensor coverage, while sub-$\mu\mathrm{T}$ mean absolute error is achieved for $B_0\le 100~\mu\mathrm{T}$ and for the 98-point training case at $B_0=400~\mu\mathrm{T}$. 
This performance supports the use of the framework for high-precision magnetic-field monitoring in constrained experimental environments, where restricted sensor access due to vacuum chambers, detector structures, beamline apertures, or mechanical constraints necessitates reconstruction from limited measurement points. However, the present experiment should not be interpreted as a demonstration of sub-ppm capability. The validation apparatus exhibits field variations of order 10\% across the measurement volume, whereas many high-precision infrastructures are engineered to provide more uniform, stable, and well-controlled magnetic environments. Greater field uniformity should reduce the complexity of the reconstruction problem, but the attainable precision will also depend on probe calibration, temporal stability, ambient-noise suppression, and systematic control. Thus, extension of the present method to sub-ppm or other ultra-high-precision applications requires dedicated validation under substantially more uniform and stable field conditions.

\section{Conclusion and Outlook}

This work presents a physics-informed neural network framework specifically engineered for high-precision three-dimensional magnetic field reconstruction in constrained experimental environments. The core advancement lies in a dual-constrained optimization strategy that enforces Maxwell's equations not only across the computational domain via random collocation but also explicitly at sensor coordinates where measurement data is most reliable. This approach ensures physical consistency at the measurement locations while maintaining divergence-free and curl-free conditions throughout the volume.

The methodology achieves exceptional performance across both synthetic and physical validation scenarios. Numerical experiments based on the Biot-Savart law demonstrate reconstruction accuracies on the order of $10^{-4}$, representing a tenfold improvement over existing PINN-based magnetic field mapping techniques. This enhancement stems directly from the explicit physics residual enforcement at measurement locations and the residual-based adaptive refinement (RAR) strategy, which dynamically concentrates computational resources on regions where physical constraints are most violated.

Experimental validation using a custom-engineered coil assembly confirms the framework's robustness under real-world conditions. 
Across the tested dynamic range, the reconstruction reaches sub-percent relative accuracy, with the best cases at the $10^{-3}$ level under ambient conditions.
This performance is maintained even with severely sparse sampling (30 surface points), validating the framework's suitability for scenarios where sensor access is restricted by vacuum chambers or complex detector geometries.

Compared to traditional spherical harmonic expansion methods, the proposed approach eliminates truncation artifacts and avoids the resolution-noise trade-off inherent in finite-order multipole expansions. The continuous functional representation enables arbitrary spatial resolution without interpolation errors, providing a significant advantage for high-gradient regions in particle physics beamlines.

These results establish the dual-constraint PINN framework as a viable high-precision tool for magnetic-field mapping in complex experimental environments. The simulation validation achieves a reconstruction accuracy of $10^{-4}$, while the physical experiment demonstrates up to $10^{-3}$ relative accuracy under ambient conditions. The PINN framework also shows robustness against sparse sampling and noise. It is important to note that the current results do not demonstrate sub-ppm capability, given the limitations of the present experimental setup. The results nevertheless provide a useful foundation for future tests in more uniform, stable, and well-controlled high-precision experimental infrastructures, where the attainable precision must be assessed together with probe calibration, temporal stability, and other systematic effects.

\begin{acknowledgments}
We thank Tsutomu Mibe, Mitsushi Abe, Shinji Ogawa and other colleagues in the J-PARC Muon $g-2$/EDM Collaboration for helpful comments and discussions.
This work was supported by National Natural Science Foundation
of China (Grants No. 12305217, 12475108).
\end{acknowledgments}

\end{document}